\documentclass[prl,aps,twocolumn,superscriptaddress]{revtex4-1}
\usepackage{bm}
\usepackage[colorlinks=true,linkcolor=blue]{hyperref}
\usepackage{amsmath}
\usepackage{amssymb}
\usepackage{amsthm}
\usepackage{amsfonts}
\usepackage{times}
\usepackage{latexsym}
\usepackage{graphicx}
\usepackage{color}

\begin{document}

\title{Superconductor to Mott insulator transition in YBa$_2$Cu$_3$O$_7$/LaCaMnO$_3$ heterostructures}

\author{B. A.\ Gray}
\affiliation{%
Department of Physics, University of Arkansas, Fayetteville, Arkansas 70701, USA}
\author{S. Middey}
\email{smiddey@uark.edu}
\affiliation{%
Department of Physics, University of Arkansas, Fayetteville, Arkansas 70701, USA}

\author{G.\ Conti}
\affiliation{%
Department of Physics, University of California, Davis, California 95616, USA}
\affiliation{%
Materials Sciences Division, Lawrence Berkeley National Laboratory, Berkeley 94720, USA}
\author{A. X.\ Gray}
\affiliation{%
Department of Physics, Temple University, Philadelpia, Pennsylvania 19130, USA}
\author{C.-T.\ Kuo}
\affiliation{%
Department of Physics, University of California, Davis, California 95616, USA}
\affiliation{%
Materials Sciences Division, Lawrence Berkeley National Laboratory, Berkeley 94720, USA}
\author{A. M. Kaiser}
\affiliation{%
Oerlikon Leybold Vacuum GmbH, K\"{o}ln, Germany}

\author{S. Ueda} 
\affiliation{%
Synchrotron X-ray Station at SPring-8, National Institute for Materials Science, Hyogo 679-5148, Japan}
\author{K. Kobayashi}
\affiliation{%
Synchrotron X-ray Station at SPring-8, National Institute for Materials Science, Hyogo 679-5148, Japan}
\author{D. Meyers}
\affiliation{%
Department of Physics, University of Arkansas, Fayetteville, Arkansas 70701, USA}
\author{M. Kareev}
\affiliation{%
Department of Physics, University of Arkansas, Fayetteville, Arkansas 70701, USA}
\author{I. C.\ Tung}
\affiliation{%
Advanced Photon Source, Argonne National Laboratory, Argonne, Illinois 60439, USA}
\affiliation{%
Department of Materials Science and Engineering, Northwestern University, Evanston, Illinois 60208, USA}
\author{Jian\ Liu}
\affiliation{%
Department of Physics and Astronomy, University of Tennessee, Knoxville, Tennessee 37996, USA}
\author{C. S.\ Fadley}
\affiliation{%
Department of Physics, University of California, Davis, California 95616, USA}
\affiliation{%
Materials Sciences Division, Lawrence Berkeley National Laboratory, Berkeley 94720, USA}
\author{J.\ Chakhalian}
\affiliation{%
Department of Physics, University of Arkansas, Fayetteville, Arkansas 70701, USA}
\affiliation{Department of Physics and Astronomy, Rutgers University, Piscataway, New Jersey 08854, USA}
\author{J. W.\ Freeland}
\email{freeland@anl.gov}
\affiliation{%
Advanced Photon Source, Argonne National Laboratory, Argonne, Illinois 60439, USA}

\begin{abstract}
The superconductor-to-insulator transition (SIT) induced by means such as external magnetic fields,  disorder  or spatial confinement is a vivid illustration of a quantum phase transition dramatically affecting the  superconducting  order parameter. In pursuit of a new realization of the SIT by interfacial charge transfer,  we developed  extremely thin superlattices composed of high $T_c$ superconductor  YBa$_2$Cu$_3$O$_7$ (YBCO) and colossal magnetoresistance ferromagnet La$_{0.67}$Ca$_{0.33}$MnO$_3$ (LCMO). By using   linearly  polarized  resonant  X-ray absorption spectroscopy and magnetic circular dichroism, combined with hard X-ray photoelectron spectroscopy, we derived  a complete picture of the interfacial carrier doping in cuprate and manganite atomic layers, leading to the transition from superconducting to an unusual Mott insulating state emerging with the increase of LCMO layer thickness. In addition, contrary to the common perception that only transition metal ions may response to the charge transfer process, we found that charge is also actively compensated by rare-earth and alkaline-earth metal ions of the interface. Such deterministic control of $T_c$ by pure electronic doping without any hindering effects of chemical substitution is another promising route to disentangle  the role of disorder on the pseudo-gap and charge density wave phases  of underdoped cuprates. 

\end{abstract}

\date{\today}
\maketitle

Heterointerfaces between  dissimilar layers of semiconductors have long been at the forefront of condensed matter physics in topics ranging from charge transfer in a  $p-n$ junction to several Hall effects arising in two-dimensional electron gases. For the case of complex oxide interfaces, the past decade has witnessed a rapid growth in science and  ability to synthesize interfaces between a wide variety of oxides with  atomic layer precision~\cite{Zubko:2011wea,Bibes:2011ko,Bhattacharya:2014ey,Chakhalian:2014ef}. Despite the marked progress, many fundamental questions still remain not well understood, particularly in regards to how charge carriers redistribute across interfaces in  strongly correlated electron systems and how this process affects spin, charge and orbital  degrees of freedom. Macroscopically, two factors seem to be of key importance: charge transfer localization at the interface and the presence of a longer scale charge transfer related to the balancing of the mismatch in chemical potential. The former has been observed at many perovskite interfaces where the local interactions drive a charge transfer. For example, in cases like La$B^{3+}$O$_3$/LaNi$^{3+}$O$_3$, the $B$ site ion and Ni form a redox couple resulting in $B^{4+}$ and Ni$^{2+}$ charges localized within a very  short range of 1-2 unit cells (u.c.) of the interface~\cite{Hoffman:2013vm,Disa:2015vo,Cao:2016dj,Grisolia:2016il}. As for  the balancing of the chemical potential via charge transfer, which is  well understood and successfully utilized for semiconductor junctions,   it is still an open research area  with many  experimental  and theoretical  challenges due to  the strongly correlated nature of the  carriers~\cite{Okamoto:2004in,Oka:2005fl,Yunoki:2007ei,Lee:2007fv,Charlebois:2013eb}.

The case of  YBa$_{2}$Cu$_{3}$O$_{7-x}$/La$_{0.67}$Ca$_{0.33}$MnO$_3$  (YBCO/LCMO) interface provides a ideal system where antagonistic combination of unconventional magnetism and high $T_c$ superconductivity defines a number of intriguing  phenomena governed by the charge  redistribution between magnetic electrons of manganite and  Cooper pairs    of  the cuprate to equalize chemical potential across the boundary (see Ref. \onlinecite{Chakhalian:2014ef} and references therein).  From the perspective of interfacial charge transfer and its  impact on the overall materials properties of the YBCO/LCMO heterojunction, there has been a significant number of results  including  charge, spin and orbital  reconstruction as revealed by  electron microscopy\cite{Salafranca:2010fg,Chien:2013cb,Salafranca:2014eq,Cuellar:2014fz,Biskup:2015hh}, neutron reflectometry~\cite{Stahn:2005km,Hoffmann:2005bj,Hoppler:2009hv}, resonant  X-rays~\cite{Chakhalian:2007fq,Werner:2010dn,Visani:2011fg,Satapathy:2012fy,Liu:2012bc,Cuellar:2014fz,Frano2016}, and cross-sectional STM~\cite{Fridman:2011gq,Chien:2013cb}, where the reported length scales of charge transfer effects range from a few\ \AA\ to  several nanometers. The variability in scale of these charge transfer effects
stems from the exploration of thick YBCO$_{7}$ and LCMO
constituent layers of multiple unit cells (u.c.) for which the bulk physics of the inner layers can obscure the
interface induced many-body behaviour. Based on this  observation, we can then ask  what  happens  when the LCMO and YBCO layers approach a 1-2 u.c. limit and the entire structure becomes  purely interfacial in nature.

\begin{figure*}
\vspace{-0pt}
\begin{tabular}{r}
     \includegraphics[width=.7\textwidth]{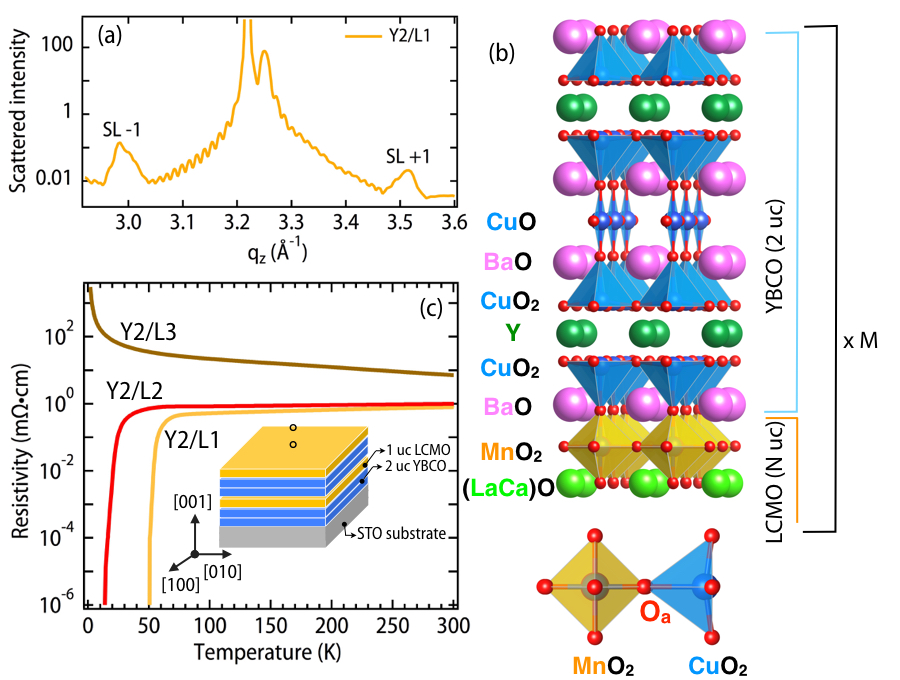} 
\end{tabular}
\caption{\label{Fig1} {\bf Crystal structure and electrical transport.} {\bf (a)} XRD scan for Y2/L1 sample showing clear thickness fringes and superlattice (SL) peaks.   {\bf (b)} The lattice structure consistent with the diffraction data of panel (a). {\bf (c)} Electrical transport measurements showing a clear superconductor to insulator transition as a function of LCMO layer thickness. Inset shows growth sequence for Y2/L1 sample.}
\end{figure*}

To address this question, we explored  a set of  interface-controlled layered heterostructures  composed of YBa$_{2}$Cu$_{3}$O$_{7}$ (2 u.c.) and La$_{0.67}$Ca$_{0.33}$MnO$_3$ (N=1, 2 and 3 u.c.)  to investigate the emergent  behaviors due to  the  reconstructed charge   and  magnetic moments   at the complex oxide interface.  By utilizing the combined power of  resonant X-ray absorption (XAS)\  and hard X-ray photoelectron spectroscopy (HXPS), we are able to elucidate how the large transfer of electrons from LCMO effectively tunes the YBCO layer across the doping phase space from being a superconductor (SC) into  the antiferromagnetic (AFM)\ phase. Additionally, the electronic responses of  Ba  and  La  atoms demonstrate  their important roles as electron acceptor and electron donor, respectively, sharing the doped charges with the CuO$_2$ plane, thus revealing a strong similarity to doping by oxygen content, $x$ in the bulk YBa$_2$Cu$_3$O$_{7-x}$.

\section*{Results}
The structure was determined by X-ray diffraction. The superlattice (SL) period of $\sim$22\ \AA\ from the spacing of the diffraction peaks (Fig.\ \ref{Fig1}(a)) for N=1 SL is consistent with the structure shown in Fig.\ \ref{Fig1}((b). This particular superlattice structure was based upon the detailed electron microscopy studies from several groups, which found  that the interfaces predominantly involve an MnO$_2$ - \textit{A}O - CuO$_2$ sequence~\cite{Zhang:2009eu,Visani:2011fg,Chien:2013cb,Zhang:2013fw}. The superlattice spacing was consistent with extracting the value from the total thickness measured by the fringes. As shown in Fig.\ \ref{Fig1}(c), electrical transport measurements demonstrate a clear superconductor-insulator (S-I) transition with increasing N, where the superconducting transition, as determined by the midpoint, is reduced from 50 K to zero and the N = 3 data is consistent with the anti-ferromagnetic insulating phase in the lightly hole doped region of the phase diagram. While the dependence of the $T_c$ seems consistent with a decrease in hole doping with increasing LCMO layer thickness, it turns out that the data has a different evolution than is observed for the case of YBa$_2$Cu$_3$O$_{7-\delta}$~\cite{Ito:1993ii,Ando:2004ww,Barisic:2013id}. The most notable is a linear $T$-dependence that disappears  quickly in the bulk as a function of $\delta$, while the resistivity in the normal state of the N = 1, 2 superlattices remains linear with temperature down to $\sim$100 K.

\begin{figure}[h]\vspace{-0pt}
\begin{tabular}{r}
     \includegraphics[width=0.45\textwidth]{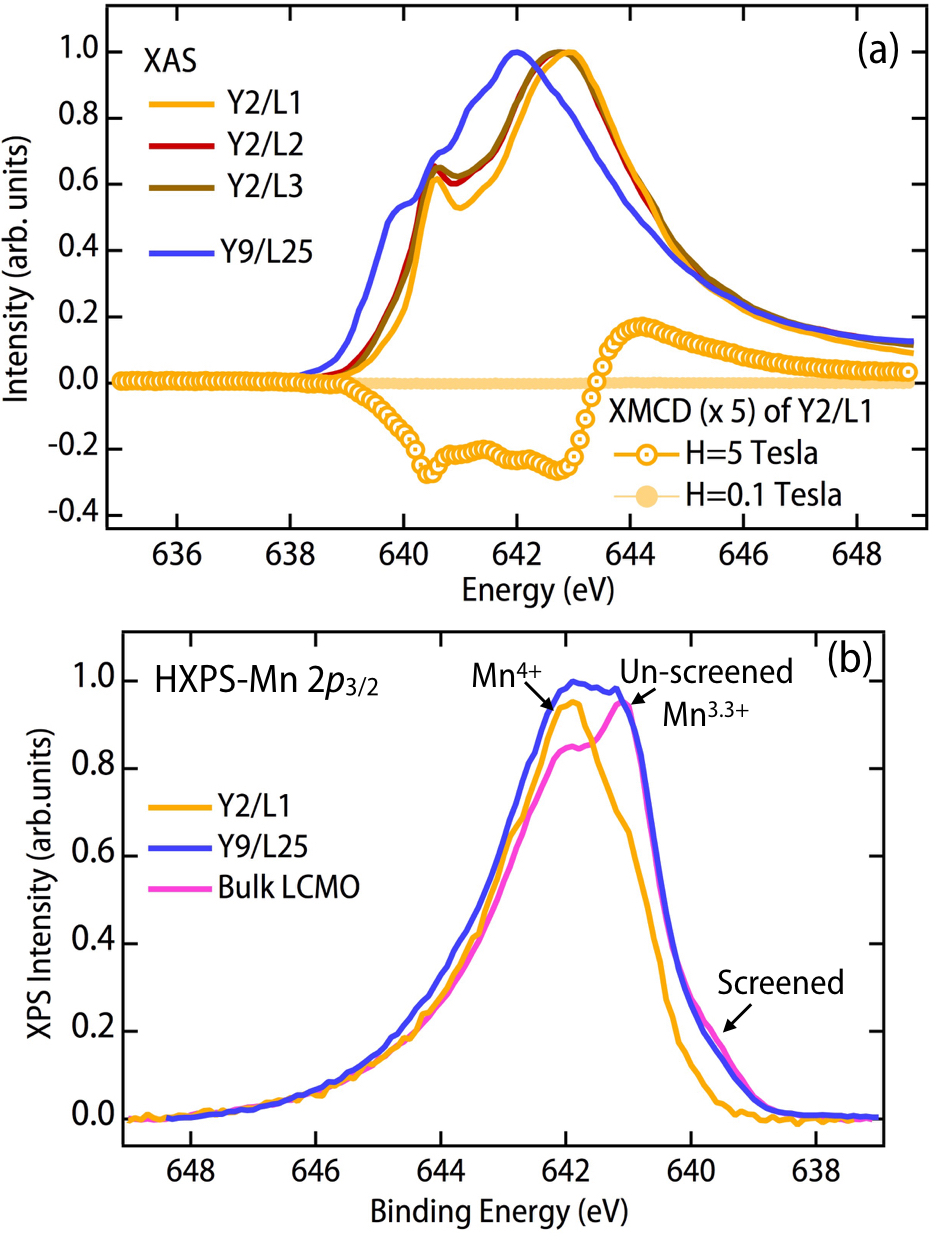} 
\end{tabular}
\caption{\label{mnxas} {\bf Heterostructuring effect on manganite layers.} {\bf (a)} Mn $L_3$-edge x-ray absorption spectroscopy demonstrating that for up to 3 unit cells of LCMO, the Mn valence is fixed close to 4$^+$ rather than the expected 3.33$^+$. For comparison, we show data from a thicker LMCO/YBCO superlattice (Y9/L25) with a nominal Mn valence of 3.33$^+$. {\bf (b)} HXPS spectra of Mn 2$p_{3/2}$, demonstrating that in Y9/L25 and bulk LCMO, Mn has nominal valence of 3.33$^+$, whereas Mn in the Y2/L1 SL has nominal valence of 4$^+$.}
\end{figure}

To understand the transition from superconductor to insulating state, we made use of polarized soft-XAS and hard-XPS to extract an element-resolved picture of the electronic and magnetic states. Beginning with the Mn $L$-edge X-ray absorption spectroscopy (XAS), we find that, rather than the expected 3.33+ valence~\cite{Schiffer:1995cj,Chmaissem:2003fx}, all the films demonstrate a Mn valence state of 4+ (see Figure\ \ref{mnxas}(a)). In the bulk phase diagram of manganites~\cite{Schiffer:1995cj}, this doping level corresponds to an insulator and the magnetism is either paramagnetic (PM) or antiferromagnetic (AFM) as noted by the small X-ray magnetic circular dichroism at 5T, which vanishes as the field tends to zero. In addition, Fig.\ \ref{mnxas}(b) shows HXPS spectra of Mn 2$p_{3/2}$ of the Y2/L1, Y9/L25, and LCMO bulk samples. With comparison to the prior HXPS studies of LCMO and LSMO~\cite{Horiba:2005bz,Mannella:2008cl}, we note the effects of final-state core-hole screening in such spectra.  For bulk LCMO, one can observe two shoulders in addition to the unscreened peak of Mn $2p_{3/2}$ at a binding energy (BE) of 641.1 eV.  The shoulder at BE = 642.2 eV is attributed to the Mn 4+ states, whereas the unscreened peak is attributed to 3.33+ states.  The other shoulder in the low BE region at 639.5 eV is assigned to well-screened states, which are associated with delocalized screening and metallic character in the 3d-based transition metal oxides~\cite{Horiba:2005bz}.  In the Y2/L1 sample, we found that the component of Mn 4+ states is enhanced and the well-screened feature is suppressed, suggesting a charge transfer from LCMO to YBCO with the LCMO layer displaying an insulating behavior, consistent with the XAS results in Fig. 2(a).  The Y9/L25 show features similar to both the LCMO bulk and to the Y2/L1 SL, so we can infer that the bulk-like part contributes to the 3.33+ valence state, and the YBCO/LCMO interface contributes to the 4+ valence state.  We measured also the Mn 3$s$ HXPS spectrum  (not shown here) of the bulk LCMO sample. It has a multiplet splitting of $\Delta E_{3s}$ = 5.2 eV that is directly related to the Mn spin moment and charge state.  A prior study of Mn 3$s$ exchange splittings in mixed-valence manganites permits concluding that the $\Delta E_{3s}$ = 5.2 eV is consistent with the Mn nominal valence of 3.33+~\cite{Galakhov:2002hf}, again qualitatively consistent with the XAS results for Y9/L25 with the thickest LCMO layer.  Spectral interference of Mn 3$s$ and Ba 4$d$ in the SLs unfortunately prevented measuring Mn 3$s$ for the superlattice samples.  From the Mn results it is clear that the LCMO layer has lost 0.67 electrons per layer (Mn$^{3.33+} \rightarrow$ Mn$^{4+}$) in these ultra-thin superlattices. If the electrons have been transferred from the LCMO into the YBCO layer, then we can search for signs of the altered YBCO charge state in the Cu XAS. 

\begin{figure}[h]\vspace{-0pt}
\begin{tabular}{r}
     \includegraphics[width=0.45\textwidth]{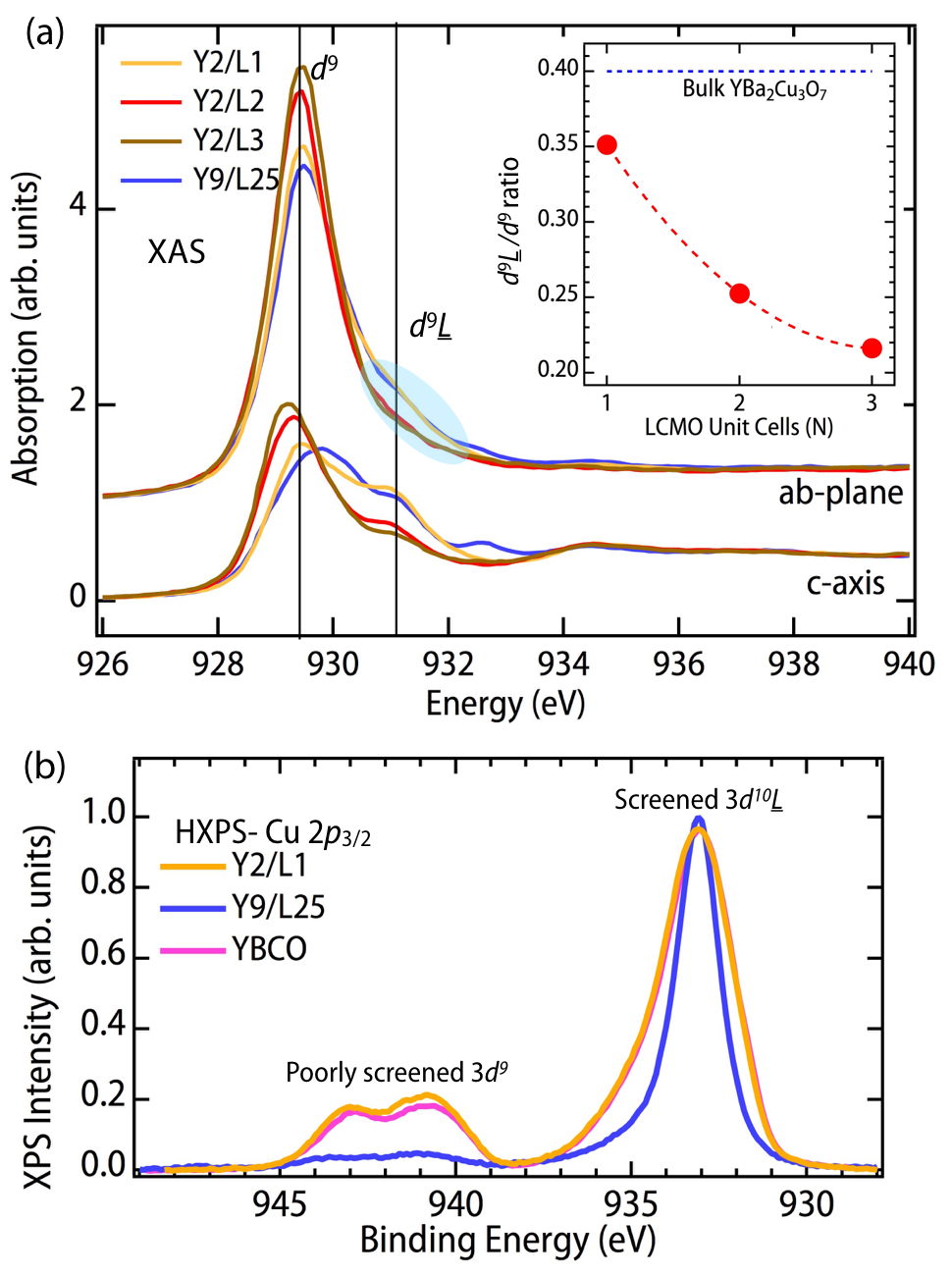} 
\end{tabular}
\caption{\label{cuxas} {\bf Heterostructuring effect on cuprate layers.}  {\bf (a)} Cu $L$-edge XAS showing the trend in the electronic structure and orbital differences with LCMO layer thickness. While all show the strong polarization dependence expected for a pure $3d_{x^2-y^2}$ state, there is a clear drop in the $3d^9\underline{L}$ state. {\bf (b)} HXPS spectra of Cu 2p$_{3/2}$ of the Y2/L1, Y9/L25 and YBCO bulk samples, where the well screened peak of Cu 2$p_{3/2}$ is observed at binding energy 932eV and corresponds to the final state $3d^{10}\underline{L}$. 
The satellite peaks at binding energy 940-944 eV correspond to the 3$d^9$ poorly screened final states. }
\end{figure}

To track the doping level of the CuO$_2$ plane, we rely on a base knowledge built upon studies of bulk YBCO~\cite{Nucker:1995hk,Merz:1998he,Hawthorn:2011db}. While the valence of Cu remains 2+ ($3d^9$) with hole doping, there is a clear emergence of a ligand hole state ($3d^9\underline{L}$) on the oxygen site, which is important for Cooper pairing~\cite{Zhang:1988jf,Meyers:2013bf}. Detailed examination of the bulk cuprates have shown that this peak on both the Cu $L$-edge and O $K$-edge track directly with doping of the CuO$_2$ plane~\cite{Peets:2009iv,Chen:2013ts}, which can be altered in  the bulk by changing oxygen stoichiometry\cite{Ito:1993ii}, replacing Y with Ca~\cite{Neumeier:1989if,ChNiedermayer:1998fu}, or doping at Cu sites~\cite{Tallon:1997gx}. Figure \ref{cuxas}(a) shows the evolution of the polarization dependent Cu $L$-edge with increasing N together with a comparison to a thicker YBCO superlattice with bulk like absorption. On this figure we note the location of the $3d^9\underline{L}$ feature associated with hole doping in the CuO$_2$ plane. Note also the absence of the feature near  932.5 eV related with the partial oxidation of the chains associated with underdoped samples~\cite{Nucker:1995hk,Merz:1998he,Hawthorn:2011db},  implying  the presence of fully
oxygenated chain layer in these superlattices.

From Fig.\ \ref{cuxas}(a)  we see that the XAS of the N=1 SL is very close to that of the thicker Y9/L25 superlattice, which has a $T_c$ of 75 K. With increasing N there is a drastic reduction in the $3d^9\underline{L}$ state and corresponding increase in the $3d^9$ feature. This evolution is directly associated with a decreasing the doping level of the CuO$_2$ plane due to the charge transfer from the LCMO layer into the YBCO layer. In order to analyze this on a more quantitative level, we have integrated the background subtracted Cu $L_3$ spectra and done a direct comparison as a function of spectral weight ratio, $3d^9\underline{L}$/$3d^9$ (see inset of Fig.\ \ref{cuxas}(a)). This ratio shows a clear drop with increasing N  consistent with a reduction of hole doping of the CuO$_2$ plane with increasing N. In comparison we also include the ratio for optimal doping as determined by an identical analysis of optimally doped YBCO from Ref. \onlinecite{Hawthorn:2011db}.  

Figure \ref{cuxas}(b) shows the HXPS spectra of Cu 2$p_{3/2}$ of the Y2/L1, Y9/L25 and YBCO bulk samples, where the well screened peak of Cu $2p_{3/2}$ is observed at BE = 932 eV and corresponds to the final state $3d^{10}\underline{L}$, implying charge transfer from oxygen due the positive core-hole potential. The satellite peaks at BE = 945-940 eV correspond to the Cu $2p3d^9$, poorly screened final state.  Previous studies indicate that the monovalent Cu$_2$O compounds show only the well screened peak and no satellites since the 3$d$ shell is already filled in the initial state.  Thus, the presence of the satellite peaks for Y2/L1 and Y9/L25 implies Cu in the 2+ state~\cite{Maiti:2009hq}.  The Cu $2p$ satellite peaks of the Y2/L1 are slightly more intense compared to those of bulk YBCO, although some of this could well be due to a small difference in the inelastic background subtraction. However, the satellite intensity of Y9/L25 is dramatically different compared to that of bulk YBCO and Y2/L1. This result indicates that the Cu valence state at the YBCO/LCMO interface is much different from the YBCO bulk-like environment.  

\begin {figure}
  \includegraphics [width=0.45\textwidth] {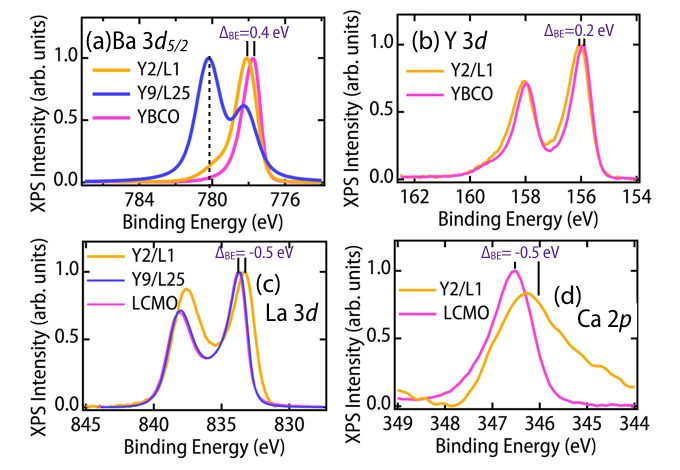}
  \caption{\label{XPS} {\bf Responses of rare-earth and alkaline earth metal elements.} Hard X-ray photoelectron spectroscopy of the core levels of the Y2/L1 sample: {\bf (a)} Ba 3$d_{5/2}$ {\bf (b)} Y 3$d$ {\bf (c)} Ca 2$p_{3/2}$ and {\bf (d)} La 3$d_{5/2}$. All of these data were recorded at $T$ = 300K and the energy was calibrated using Au.}
\end{figure}

From the behavior of the XAS data it is evident that  the amount of electrons doped to Cu sites is much smaller than the 0.67 x N electrons  donated by the LCMO block. In order to find out how the charge transfer influences energy levels, we consider HXPS  results from other atoms in the SLs in Fig.\ \ref{XPS}, beginning with the YBCO side of the interface.  Fig.\ \ref{XPS}(a) indicates that not only the Cu $2p$ features are different in the spectrum of Y9/L25 compared to that of Y2/L1, but also the Ba $3d$ core level. In addition to the main peak of Ba $3d_{5/2}$ at BE = 778 eV, there is a strong peak at BE = 780 eV which appears only as a shoulder in the spectrum of the Y2/L1 SL, and is essentially absent in bulk YBCO.  The two-peaked structure in Y9/L25 we assign to the existence of two kinds of Ba in the superlattice structure, at the interface and away from it, consistent with the structure shown in the schematic of Fig.\ \ref{Fig1}(b). In the case of the thick LCMO overlayer for the Y9/L25 sample, the interface sensitivity is enhanced due to the short inelastic mean free path (IMFP)~\cite{Tanuma:2011ct} giving rise to a large interface peak. This feature at BE=780 eV, due to the interfacial sites, has a non-bulk-like electronic configuration.  The Ba $3d_{5/2}$ core level of Y2/L1 SL has also been shifted towards higher binding energy by $\sim$0.4 eV as compared to the bulk YBCO film. The response of this core level to the depression of $T_c$, is similar to the chemical doping effect observed in bulk YBCO$_{7-x}$ and emphasizes that the Ba-O plane also acts as an electron reservoir~\cite{Yang:1990bi}.  The Y atoms, which are farther from the interface, displayed in Fig.\ \ref{XPS}(b), only a small shift of $\sim$0.2 eV to higher binding energy and no significant change in spectral shape, which is consistent with the bulk doping behavior as well~\cite{Yang:1990bi,Yang:1991jt}, and the fact that Y maintains a bulk-like coordination in the superlattice geometry.

On the other side of interface, the La $3d_{5/2}$ BE of Y2/L1 moves to lower binding energy by about 0.5 eV compared to Y9/25 and bulk LCMO. A BE shift to lower energy of $\sim$0.5 eV is also observed for the case of the Ca $2p_{3/2}$ XPS data as shown in Fig.\ \ref{XPS}(d), with an additional broadening toward the low-BE side, that might indicate a greater degree of interdiffusion by Ca at the interface.  A self-consistent picture of the BE shifts in Fig.\ \ref{XPS} is also possible through a measurement of the valence-band offset (VBO) at the Y2/L1 interface using HXPS core-level and valence-band maximum (VBM) energies for Y2/L1 and the bulk reference samples of YBCO and LCMO, and a method introduced into oxide studies by Chambers et al.~\cite{Chambers:2000fo,Chambers:2004fd}.  The VBO obtained in this way is 0.67 eV, with sign such that BEs in YBCO should increase and those in LCMO should decrease, just as seen in Fig.\ \ref{XPS}.  The sign of the VBO is also consistent with the effective n doping of YBCO and p doping of LCMO.  The fact that the magnitude of this VBO is of the same general magnitude as the shifts in Fig.\ \ref{XPS} nicely supports this interpretation. From the behavior of the Mn and Cu XAS and HXPS data of the Y2/LN SLs, it is thus evident that there is charge transfer from the LCMO layer into the YBCO layer, with binding energy shifts in opposite directions on the two sides.

From a chemical structure standpoint, one can consider how balancing charge in a larger unit cell helps to understand the charge transfer. The key is the {\it missing chains} at the interface that lead to a partial and charge imbalanced YBCO unit cell that needs to be compensated by the LCMO layers~\cite{Chien:2013cb}, which leads to the structure shown in Fig.~\ref{Fig1}(b). By constructing a larger unit cell composed of LCMO and YBCO unit cells together and counting the charges using the ionic values explains the direction of charge motion, but the change of the charge in the CuO$_2$ layers comes out much larger than experimentally seen. This is because we are assuming that the other elements are passive, but as shown above the other ions in the lattice are participating in the charge compensation. The shift to higher energy on the YBCO side  is consistent with the electron doping while the opposite shift on the La and Ca is due to electron depletion of the LCMO layer. The fact that Y reacts less than Ba is potentially related to the distance from the interface and suggests that within 1-2 CuO$_2$ layers that the YBCO electronic structure is returning to bulk-like properties, which is consistent with a recent study of the electronic interface with cross-sectional STM~\cite{Chien:2013cb}. This length scale is also consistent with theoretical calculations that explore how the electronic structure evolves across the LCMO/YBCO interface~\cite{Yunoki:2007ei,Pavlenko:2007bv,Pavlenko:2009kt,Salafranca:2010fga,Salafranca:2014eq,yang2009}. Based upon simple arguments in the level of the chemical potential~\cite{Yunoki:2007ei}, one can determine the expected trends that are consistent with that seen here in the experimental data.

\begin{figure}
\vspace{-0pt}
\begin{tabular}{r}
     \includegraphics[width=0.45\textwidth]{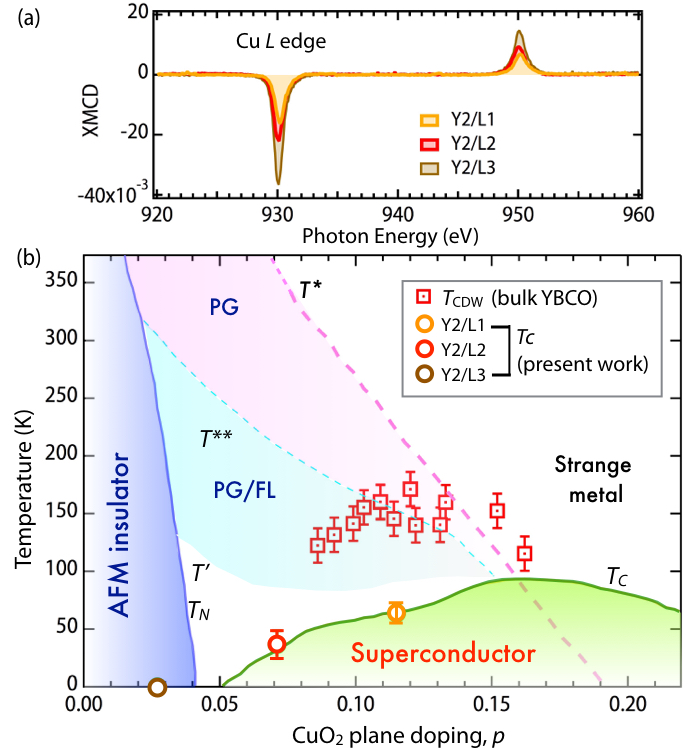} 
\end{tabular}
\caption{\label{phasediagram} {\bf Phase diagram.} {\bf(a)} XMCD of Cu $L_{3,2}$ edges for Y2/LN SLs. {\bf(b)} Doping level inferred from both $T_c$ and the XAS analysis of Y2/LN SLs are overlayed with the bulk phase diagram for YBCO, adapted from the supplemental of ref.~\onlinecite{Tabis:2014kb}. PG=pseudogap, FL= Fermi liquid. Superconducting $T_c$ is defined as the midpoint of the transition and the width of transition has been marked as the corresponding error bar.  %(T$_c$ data from ref.~\onlinecite{Liang:2006hr}).
}
\end{figure}

Taking all the experimental results together, we can confirm that the suppression of the SC state is due to doping of electrons into YBCO from the LCMO layer. To quantify this, using the $T_c$ data from the films together with data from the bulk~\cite{Liang:2006hr}, we can estimate that each LCMO layer is doping 0.67 electrons to the whole of the YBCO layer, but only $\sim$0.05 electrons into the CuO$_2$ planes as shown in Fig.\ \ref{phasediagram}. This discrepancy is perhaps not surprising though since in the bulk changing from YBa$_{2}$Cu$_{3}$O$_{7}$ to YBa$_{2}$Cu$_{3}$O$_{6}$, which is a total charge change of 2-, only leads to a small doping change in the CuO$_2$ plane~\cite{Liang:2006hr}.  A similar effect was seen with electrostatic doping experiments, which explained why so little variation of $T_c$ was found in ferroelectric/cuprate heterostructures~\cite{Salluzzo:2008dn}. This also highlights  that it is important to track all of the potential charge reservoirs in order to create a physical picture of the process in a multicomponent system.

Together with this reduction in $T_c$, we also observed a significant increase in the Cu XMCD at high-fields (5 T) connected with increase in magnetic order associated with the onset of the AFM phase. The same trend using XMCD to probe Cu magnetism was recently seen for bulk YBCO as a function of doping~\cite{DeLuca:2010iw}, which serves to support the idea that while we are doping via an interface into an ultra YBCO layer, much of the expected bulk physics remains intact. This results from the highly two-dimensional character of the cuprate high temperature superconductors, as was seen by the presence of superconductivity even in isolated CuO$_2$ layers~\cite{GLogvenov:2009eh,DiCastro:2015ux}.

Given the  striking similarity  between  the effect of chemical doping and the interfacial   ``doping" by electron/hole transfer, there are a number of interesting  questions which arise in the view of the  recent discovery  of a  charge density wave (CDW) phase competing with SC pairing. Specifically, recent extensive work on underdoped YBCO revealed the presence of long-range charge fluctuations in CuO$_2$ in a wide range of hole doping~\cite{Wu:2011ke,Chang:2012dna,Ghiringhelli:2012bw,LeTacon:2013es,Tabis:2014kb,Comin:2015ca,Comin:2016}, which act  to  suppress the SC order.  As demonstrated  above and summarized in Fig.\ \ref{phasediagram}(b), the charge transfer between manganite and cuprate layers in these superlattices brings the cuprate component the under doped region of bulk YBCO phase diagram, where CDW phase would also appear. Based on this observation, it can be anticipated that similar charge   fluctuations  in  CuO$_2$  planes are also present in the case of heterojunctions. However, the situation is not that  straightforward considering the fact that the CuO chains remain fully oxygenated in these SLs contrary to the chains with ordered oxygen vacancy of underdoped bulk YBCO. Moreover, the incipient CDW fluctuations may be strongly  altered or even entirely suppressed because of the charge mismatch across the  YBCO/LCMO interface~\cite{Chakhalian:2014ef,Frano2016,He2016}   leading to  a  possible new phase state without  CDW (e.g. charge glass) for the underdoped  part of cuprate phase diagram~\cite{Keimer:2015bn}. On the manganite side of the  junction the charge transfer causes massive   hole doping into  the CMR\ layers. This in turn raises an interesting  question  if the hallmark CMR phenomena (e.g. electronic phase separation and  Jahn-Teller like distortions from strong electron-lattice coupling) persist after pure hole doping into  intact lattice by   random chemical  doping~\cite{Tokura:2006ff}.

In conclusion, by variable-polarization XAS, MCD, and HXPS, we have shown a clear superconductor to insulator transition driven by interfacial doping in an ultra-thin cuprate-maganite superlattice. Without changing the concentration of vacancies in the chains, we can utilize the chemical potential imbalance at the LCMO/YBCO interface to dope electrons into YBCO over a wide range of the phase diagram. This offers an interesting route to examine the link between dopant disorder and phenomena in underdoped superconductors. As noted in recent work, there is a clear correlation between the spatial dopant arrangements and charge order observed in the cuprates~\cite{Ricci:2013hm,Campi:2015cv}. Given the dramatic differences associated with how the ordering of dopants can affect the corresponding phase diagrams both in bulk~\cite{Akahoshi:2003hk} and heterostructures~\cite{Santos:2011tj,NelsonCheeseman:2014gm}, this might suggest a modified phase diagram for the YBCO doped in this manner. As noted above, the deviation from linear $T$, which is associated with the onset of the pseudogap and associated {\bf q}=0 magnetic order in the bulk~\cite{Li:2008dy,Tabis:2014kb}, occurs at a much lower temperature and perhaps suggests a suppression of the onset of the pseudogap phase. Future work, will explore further how this type of doping into ultrathin YBCO affects the phase diagram.

\subsection*{Methods}
The high-quality epitaxial superlattices (SL) consisting of 2 unit cell (uc)  of YBa$_{2}$Cu$_{3}$O$_{7}$ (YBCO) and N uc of La$_{0.67}$Ca$_{0.33}$MnO$_3$ (LCMO)  (N = 1, 2, 3) were grown by pulsed laser interval deposition on 5$\times$5~mm$^2$ SrTiO$_{3}$ (001) substrate. The layer-by-layer growth these 20 repeat superlattices (labelled as Y2/LN with N=1,2,3) were monitored by in-situ reflection high energy electron diffraction (RHEED). In addition to these superlattices, a thick bulk-like LCMO, YBCO, and (Y9/L25)$_3$ were also grown as a reference. Samples were capped with a protective coating of 4 u.c. of SrTiO$_3$, which is very stable in air.
 
The SL structures were studied by X-ray diffraction (XRD) in air at room temperature using a standard four-circle diffractometer, operating at Beamline 33-BM-C of the Advanced Photon Source (APS). The dc transport properties were measured from 300 to 2K in a Physical Property Measurement System (PPMS, Quantum Design) using the van der Pauw geometry. The XAS measurements (X-ray Linear Dichroism measurements (XLD) and X-ray Magnetic Circular Dichroism (XMCD)) were performed at the Cu $L_3$ edge and Mn $L_3$ edge, respectively, at the 4-ID-C beam line of the APS. The XLD spectra were measured in bulk sensitive total-fluorescence-yield mode, and energy calibrations were carried out by measuring a CuO (Cu$^{2+}$) standard simultaneously in the diagnostic section of the beamline. Each spectrum was normalized to the beam intensity monitored by a gold mesh set in front of the samples. In order to avoid spurious XMCD signals, XMCD spectra were recorded with both $\pm$5 Tesla magnetic fields.
 
  (HXPS) data were obtained at beamline 15XU of the SPring-8 Synchrotron. (Y2/L1) SL and the bulk YBCO and LCMO samples were measured at a photon energy of  3.238 keV and an overall energy resolution of 0.25 eV.  A Y9/L25 SL, with the structure from top to bottom of 4 u.c. STO/[25 u.c. LCMO/9 u.c. YBCO]x3/STO substrate, was measured at a photon energy of 2.2 keV and a resolution of 0.1 eV. Using these multi-keV photon energies yields more buried interface sensitivity than normal soft x-ray photoemission, through increased mean depths of emission, as controlled by the photoelectron inelastic mean free path (IMFP): as estimated from the TPP-2M formula~\cite{Tanuma:2011ct}, the IMFP at 3.238 keV will be  $\sim$ 46\ \AA, and at 2.2 keV  $\sim$ 33\ \AA.  Thus, our HXPS measurements are sensitive to approximately the first two bilayers in the Y2/LN SL, and to the bulk behavior in the thicker reference samples.  The binding energies of the HXPS spectra were calibrated using Au 4$f$ and Au $E_F$ before and after each data acquisition. The experiment was performed at room temperature and at $T$ = 20 K, below the normal superconducting  $T_C$ for YBCO; however, it is noteworthy that cooling to the superconducting state did not  induce any significant changes in any of the SL spectra, and the data reported here are thus only those collected at room temperature.

%\end{thebibliography}

%\bibliography{YBCOLCMOInt_Bib}

\begin{thebibliography}{10}
\expandafter\ifx\csname url\endcsname\relax
  \def\url#1{\texttt{#1}}\fi
\expandafter\ifx\csname urlprefix\endcsname\relax\def\urlprefix{URL }\fi
\providecommand{\bibinfo}[2]{#2}
\providecommand{\eprint}[2][]{\url{#2}}

\bibitem{Zubko:2011wea}
\bibinfo{author}{Zubko, P.}, \bibinfo{author}{Gariglio, S.},
  \bibinfo{author}{Gabay, M.}, \bibinfo{author}{Ghosez, P.} \&
  \bibinfo{author}{Triscone, J.-M.}
\newblock \bibinfo{title}{{Interface Physics in Complex Oxide
  Heterostructures}}.
\newblock \emph{\bibinfo{journal}{Annual Review Of Condensed Matter Physics}} \textbf{\bibinfo{volume}{2}}, \bibinfo{pages}{141}
  (\bibinfo{year}{2011}).

\bibitem{Bibes:2011ko}
\bibinfo{author}{Bibes, M.}, \bibinfo{author}{Villegas, J.~E.} \&
  \bibinfo{author}{Barthelemy, A.}
\newblock \bibinfo{title}{{Ultrathin oxide films and interfaces for electronics
  and spintronics}}.
\newblock \emph{\bibinfo{journal}{Advances in Physics}}
  \textbf{\bibinfo{volume}{60}}, \bibinfo{pages}{5--84} (\bibinfo{year}{2011}).

\bibitem{Bhattacharya:2014ey}
\bibinfo{author}{Bhattacharya, A.} \& \bibinfo{author}{May, S.~J.}
\newblock \bibinfo{title}{{Magnetic Oxide Heterostructures}}.
\newblock \emph{\bibinfo{journal}{Annual Review Of Materials Research}}
  \textbf{\bibinfo{volume}{44}}, \bibinfo{pages}{65--90}
  (\bibinfo{year}{2014}).

\bibitem{Chakhalian:2014ef}
\bibinfo{author}{Chakhalian, J.}, \bibinfo{author}{Freeland, J.~W.},
  \bibinfo{author}{Millis, A.~J.}, \bibinfo{author}{Panagopoulos, C.} \&
  \bibinfo{author}{Rondinelli, J.~M.}
\newblock \bibinfo{title}{{Colloquium: Emergent properties in plane
  view: Strong correlations at oxide interfaces}}.
\newblock \emph{\bibinfo{journal}{Rev. Mod. Phys.}}
  \textbf{\bibinfo{volume}{86}}, \bibinfo{pages}{1189--1202}
  (\bibinfo{year}{2014}).

\bibitem{Hoffman:2013vm}
\bibinfo{author}{Hoffman, J.} \emph{et~al.}
\newblock \bibinfo{title}{{Charge transfer and interfacial magnetism in
  (LaNiO3)n/(LaMnO3)2 superlattices}}.
\newblock \emph{\bibinfo{journal}{Phys. Rev. B}}
  \textbf{\bibinfo{volume}{88}}, \bibinfo{pages}{144411}
  (\bibinfo{year}{2013}).

\bibitem{Disa:2015vo}
\bibinfo{author}{Disa, A.~S.} \emph{et~al.}
\newblock \bibinfo{title}{{Orbital Engineering in Symmetry-Breaking Polar
  Heterostructures}}.
\newblock \emph{\bibinfo{journal}{Phys. Rev. Lett.}}
  \textbf{\bibinfo{volume}{114}}, \bibinfo{pages}{026801}
  (\bibinfo{year}{2015}).

\bibitem{Cao:2016dj}
\bibinfo{author}{Cao, Y.} \emph{et~al.}
\newblock \bibinfo{title}{{Engineered Mott ground state in a
  LaTiO$_{3+\delta}$/LaNiO$_3$ heterostructure}}.
\newblock \emph{\bibinfo{journal}{Nature Communications}}
  \textbf{\bibinfo{volume}{7}}, \bibinfo{pages}{10418} (\bibinfo{year}{2016}).

\bibitem{Grisolia:2016il}
\bibinfo{author}{Grisolia, M.~N.} \emph{et~al.}
\newblock \bibinfo{title}{{Hybridization-controlled charge transfer and induced
  magnetism at correlated oxide interfaces}}.
\newblock \emph{\bibinfo{journal}{Nature Physics}}
  \textbf{\bibinfo{volume}{12}}, \bibinfo{pages}{484--492}
  (\bibinfo{year}{2016}).

\bibitem{Okamoto:2004in}
\bibinfo{author}{Okamoto, S.} \& \bibinfo{author}{Millis, A.~J.}
\newblock \bibinfo{title}{{Electronic reconstruction at an interface between a
  Mott insulator and a band insulator}}.
\newblock \emph{\bibinfo{journal}{Nature}} \textbf{\bibinfo{volume}{428}},
  \bibinfo{pages}{630} (\bibinfo{year}{2004}).

\bibitem{Oka:2005fl}
\bibinfo{author}{Oka, T.} \& \bibinfo{author}{Nagaosa, N.}
\newblock \bibinfo{title}{{Interfaces of Correlated Electron Systems: Proposed
  Mechanism for Colossal Electroresistance}}.
\newblock \emph{\bibinfo{journal}{Phys. Rev. Lett.}}
  \textbf{\bibinfo{volume}{95}}, \bibinfo{pages}{266403}
  (\bibinfo{year}{2005}).

\bibitem{Yunoki:2007ei}
\bibinfo{author}{Yunoki, S.} \emph{et~al.}
\newblock \bibinfo{title}{{Electron doping of cuprates via interfaces with
  manganites}}.
\newblock \emph{\bibinfo{journal}{Phys. Rev. B}}
  \textbf{\bibinfo{volume}{76}}, \bibinfo{pages}{064532}
  (\bibinfo{year}{2007}).

\bibitem{Lee:2007fv}
\bibinfo{author}{Lee, W.-C.} \& \bibinfo{author}{Macdonald, A.}
\newblock \bibinfo{title}{{Electronic interface reconstruction at
  polar-nonpolar Mott-insulator heterojunctions}}.
\newblock \emph{\bibinfo{journal}{Phys. Rev. B}}
  \textbf{\bibinfo{volume}{76}}, \bibinfo{pages}{075339}
  (\bibinfo{year}{2007}).

\bibitem{Charlebois:2013eb}
\bibinfo{author}{Charlebois, M.}, \bibinfo{author}{Hassan, S.~R.},
  \bibinfo{author}{Karan, R.}, \bibinfo{author}{S{\'e}n{\'e}chal, D.} \&
  \bibinfo{author}{Tremblay, A.-M.~S.}
\newblock \bibinfo{title}{{Mott p-n junctions in layered materials}}.
\newblock \emph{\bibinfo{journal}{Phys. Rev. B}}
  \textbf{\bibinfo{volume}{87}}, \bibinfo{pages}{035137}
  (\bibinfo{year}{2013}).

\bibitem{Salafranca:2010fg}
\bibinfo{author}{Salafranca, J.} \& \bibinfo{author}{Okamoto, S.}
\newblock \bibinfo{title}{{Unconventional Proximity Effect and Inverse
  Spin-Switch Behavior in a Model Manganite-Cuprate-Manganite Trilayer
  System}}.
\newblock \emph{\bibinfo{journal}{Phys. Rev. Lett.}}
  \textbf{\bibinfo{volume}{105}}, \bibinfo{pages}{256804}
  (\bibinfo{year}{2010}).

\bibitem{Chien:2013cb}
\bibinfo{author}{Chien, T.~Y.} \emph{et~al.}
\newblock \bibinfo{title}{{Visualizing short-range charge transfer at the
  interfaces between ferromagnetic and superconducting oxides}}.
\newblock \emph{\bibinfo{journal}{Nature Communications}}
  \textbf{\bibinfo{volume}{4}} (\bibinfo{year}{2013}).

\bibitem{Salafranca:2014eq}
\bibinfo{author}{Salafranca, J.} \emph{et~al.}
\newblock \bibinfo{title}{{Competition between Covalent Bonding and Charge
  Transfer at Complex-Oxide Interfaces}}.
\newblock \emph{\bibinfo{journal}{Phys. Rev. Lett.}}
  \textbf{\bibinfo{volume}{112}}, \bibinfo{pages}{196802}
  (\bibinfo{year}{2014}).

\bibitem{Cuellar:2014fz}
\bibinfo{author}{Cuellar, F.~A.} \emph{et~al.}
\newblock \bibinfo{title}{{Reversible electric-field control of magnetization
  at oxide interfaces}}.
\newblock \emph{\bibinfo{journal}{Nature Communications}}
  \textbf{\bibinfo{volume}{5}} (\bibinfo{year}{2014}).

\bibitem{Biskup:2015hh}
\bibinfo{author}{Bi{\v s}kup, N.}, \bibinfo{author}{Das, S.},
  \bibinfo{author}{Gonzalez-Calbet, J.~M.}, \bibinfo{author}{Bernhard, C.} \&
  \bibinfo{author}{Varela, M.}
\newblock \bibinfo{title}{{Atomic-resolution studies of epitaxial strain
  release mechanisms in La1.85Sr0.15CuO4/La0.67Ca0.33MnO3superlattices}}.
\newblock \emph{\bibinfo{journal}{Phys. Rev. B}}
  \textbf{\bibinfo{volume}{91}}, \bibinfo{pages}{205132}
  (\bibinfo{year}{2015}).

\bibitem{Stahn:2005km}
\bibinfo{author}{Stahn, J.} \emph{et~al.}
\newblock \bibinfo{title}{{Magnetic proximity effect in perovskite
  superconductor/ferromagnet multilayers}}.
\newblock \emph{\bibinfo{journal}{Phys. Rev. B}}
  \textbf{\bibinfo{volume}{71}}, \bibinfo{pages}{140509}
  (\bibinfo{year}{2005}).

\bibitem{Hoffmann:2005bj}
\bibinfo{author}{Hoffmann, A.} \emph{et~al.}
\newblock \bibinfo{title}{{Suppressed magnetization in
  La$_{0.7}$Ca$_{0.3}$MnO$_3$$\slash$YBa$_2$Cu$_3$O$_{7-\delta}$ superlattices}}.
\newblock \emph{\bibinfo{journal}{Phys. Rev. B}}
  \textbf{\bibinfo{volume}{72}}, \bibinfo{pages}{140407}
  (\bibinfo{year}{2005}).

\bibitem{Hoppler:2009hv}
\bibinfo{author}{Hoppler, J.} \emph{et~al.}
\newblock \bibinfo{title}{{Giant superconductivity-induced modulation of the
  ferromagnetic magnetization in a cuprate-manganite superlattice}}.
\newblock \emph{\bibinfo{journal}{Nature Materials}}
  \textbf{\bibinfo{volume}{8}}, \bibinfo{pages}{315--319}
  (\bibinfo{year}{2009}).

\bibitem{Chakhalian:2007fq}
\bibinfo{author}{Chakhalian, J.} \emph{et~al.}
\newblock \bibinfo{title}{{Orbital reconstruction and covalent bonding at an
  oxide interface}}.
\newblock \emph{\bibinfo{journal}{Science}} \textbf{\bibinfo{volume}{318}},
  \bibinfo{pages}{1114--1117} (\bibinfo{year}{2007}).

\bibitem{Werner:2010dn}
\bibinfo{author}{Werner, R.} \emph{et~al.}
\newblock \bibinfo{title}{{YBa$_2$Cu$_3$O$_7$/La$_{0.7}$Ca$_{0.3}$MnO$_3$ bilayers: Interface
  coupling and electric transport properties}}.
\newblock \emph{\bibinfo{journal}{Phys. Rev. B}}
  \textbf{\bibinfo{volume}{82}}, \bibinfo{pages}{224509}
  (\bibinfo{year}{2010}).

\bibitem{Visani:2011fg}
\bibinfo{author}{Visani, C.} \emph{et~al.}
\newblock \bibinfo{title}{{Symmetrical interfacial reconstruction and magnetism
  in La$_{0.7}$Ca$_{0.3}$MnO$_{3}$/YBa$_{2}$Cu$_{3}$O$_{7}$/La$_{0.7}$Ca$_{0.3}$Mn$O_{3}$
  heterostructures}}.
\newblock \emph{\bibinfo{journal}{Phys. Rev. B}}
  \textbf{\bibinfo{volume}{84}}, \bibinfo{pages}{060405}
  (\bibinfo{year}{2011}).

\bibitem{Satapathy:2012fy}
\bibinfo{author}{Satapathy, D.} \emph{et~al.}
\newblock \bibinfo{title}{{Magnetic Proximity Effect in
  YBa$_{2}$Cu$_{3}$O$_{7}$/La$_{2/3}$Ca$_{1/3}$Mn$O_{3}$ and
  YBa$_{2}$Cu$_{3}$O$_{7}$/LaMnO$_{3+\delta}$ Superlattices}}.
\newblock \emph{\bibinfo{journal}{Phys. Rev. Lett.}}
  \textbf{\bibinfo{volume}{108}}, \bibinfo{pages}{197201}
  (\bibinfo{year}{2012}).

\bibitem{Liu:2012bc}
\bibinfo{author}{Liu, Y.} \emph{et~al.}
\newblock \bibinfo{title}{{Effect of Interface-Induced Exchange Fields on
  Cuprate-Manganite Spin Switches}}.
\newblock \emph{\bibinfo{journal}{Phys. Rev. Lett.}}
  \textbf{\bibinfo{volume}{108}}, \bibinfo{pages}{207205}
  (\bibinfo{year}{2012}).

\bibitem{Frano2016}
\bibinfo{author}{Frano, A.} \emph{et~al.}
\newblock \bibinfo{title}{{Long-range charge-density-wave proximity effect at cuprate/manganate interfaces}}.
\newblock \emph{\bibinfo{journal}{Nature Materials}}
  \textbf{\bibinfo{volume}{15}}, \bibinfo{pages}{831-834}
  (\bibinfo{year}{2016}).
  
\bibitem{Fridman:2011gq}
\bibinfo{author}{Fridman, I.}, \bibinfo{author}{Gunawan, L.},
  \bibinfo{author}{Botton, G.~A.} \& \bibinfo{author}{Wei, J. Y.~T.}
\newblock \bibinfo{title}{{Scanning tunneling spectroscopy study of c-axis
  proximity effect in epitaxial bilayer manganite/cuprate thin films}}.
\newblock \emph{\bibinfo{journal}{Phy. Rev. B}}
  \textbf{\bibinfo{volume}{84}}, \bibinfo{pages}{104522}
  (\bibinfo{year}{2011}).


  

\bibitem{Zhang:2009eu}
\bibinfo{author}{Zhang, Z.~L.}, \bibinfo{author}{Kaiser, U.},
  \bibinfo{author}{Soltan, S.}, \bibinfo{author}{Habermeier, H.-U.} \&
  \bibinfo{author}{Keimer, B.}
\newblock \bibinfo{title}{{Magnetic properties and atomic structure of La$_{2/3}$Ca$_{1/3}$MnO$_3$/YBa$_2$Cu$_3$O$_7$
  heterointerfaces}}.
\newblock \emph{\bibinfo{journal}{Appl. Phys. Lett.}}
  \textbf{\bibinfo{volume}{95}}, \bibinfo{pages}{242505}
  (\bibinfo{year}{2009}).

\bibitem{Zhang:2013fw}
\bibinfo{author}{Zhang, H.}, \bibinfo{author}{Gauquelin, N.},
  \bibinfo{author}{Botton, G.~A.} \& \bibinfo{author}{Wei, J. Y.~T.}
\newblock \bibinfo{title}{{Attenuation of superconductivity in
  manganite/cuprate heterostructures by epitaxially-induced CuO intergrowths}}.
\newblock \emph{\bibinfo{journal}{Appl. Phys. Lett.}}
  \textbf{\bibinfo{volume}{103}}, \bibinfo{pages}{052606}
  (\bibinfo{year}{2013}).

\bibitem{Ito:1993ii}
\bibinfo{author}{Ito, T.}, \bibinfo{author}{Takenaka, K.} \&
  \bibinfo{author}{Uchida, S.}
\newblock \bibinfo{title}{{Systematic deviation from T-linear behavior in the
  in-plane resistivity of YBa$_2$Cu$_3$O$_{7-y}$: Evidence for dominant spin scattering}}.
\newblock \emph{\bibinfo{journal}{Phys. Rev. Lett.}}
  \textbf{\bibinfo{volume}{70}}, \bibinfo{pages}{3995--3998}
  (\bibinfo{year}{1993}).

\bibitem{Ando:2004ww}
\bibinfo{author}{Ando, Y.}, \bibinfo{author}{Komiya, S.},
  \bibinfo{author}{Segawa, K.}, \bibinfo{author}{Ono, S.} \&
  \bibinfo{author}{Kurita, Y.}
\newblock \bibinfo{title}{{Electronic Phase Diagram of High-Tc Cuprate
  Superconductors from a Mapping of the In-Plane Resistivity Curvature}}.
\newblock \emph{\bibinfo{journal}{Phys. Rev. Lett.}}
  \textbf{\bibinfo{volume}{93}}, \bibinfo{pages}{267001}
  (\bibinfo{year}{2004}).

\bibitem{Barisic:2013id}
\bibinfo{author}{Bari{\v s}i{\'c}, N.} \emph{et~al.}
\newblock \bibinfo{title}{{Universal sheet resistance and revised phase diagram
  of the cuprate high-temperature superconductors}}.
\newblock \emph{\bibinfo{journal}{Proceedings of the National Academy of
  Sciences}} \textbf{\bibinfo{volume}{110}}, \bibinfo{pages}{12235--12240}
  (\bibinfo{year}{2013}).

\bibitem{Schiffer:1995cj}
\bibinfo{author}{Schiffer, P.}, \bibinfo{author}{Ramirez, A.~P.},
  \bibinfo{author}{Bao, W.} \& \bibinfo{author}{Cheong, S.-W.}
\newblock \bibinfo{title}{{Low Temperature Magnetoresistance and the Magnetic
  Phase Diagram of La$_{1-x}$Ca$_{x}$MnO$_{3}$}}.
\newblock \emph{\bibinfo{journal}{Phys. Rev. Lett.}}
  \textbf{\bibinfo{volume}{75}}, \bibinfo{pages}{3336--3339}
  (\bibinfo{year}{1995}).

\bibitem{Chmaissem:2003fx}
\bibinfo{author}{Chmaissem, O.} \emph{et~al.}
\newblock \bibinfo{title}{{Structural and magnetic phase diagrams of
  La$_{1-x}$Sr$_{x}$MnO$_{3}$ and Pr$_{1-y}$Sr$_y$MnO$_3$}}.
\newblock \emph{\bibinfo{journal}{Phys. Rev. B}}
  \textbf{\bibinfo{volume}{67}}, \bibinfo{pages}{094431}
  (\bibinfo{year}{2003}).

\bibitem{Horiba:2005bz}
\bibinfo{author}{Horiba, K.} \emph{et~al.}
\newblock \bibinfo{title}{{In vacuo photoemission study of atomically
  controlled La$_{1-x}$Sr$_x$MnO$_3$ thin films: Composition dependence of the
  electronic structure}}.
\newblock \emph{\bibinfo{journal}{Phys. Rev. B}}
  \textbf{\bibinfo{volume}{71}}, \bibinfo{pages}{155420}
  (\bibinfo{year}{2005}).

\bibitem{Mannella:2008cl}
\bibinfo{author}{Mannella, N.} \emph{et~al.}
\newblock \bibinfo{title}{{Temperature-dependent evolution of the electronic
  and local atomic structure in the cubic colossal magnetoresistive manganite
  La$_{1-x}$Sr$_x$MnO$_3$}}.
\newblock \emph{\bibinfo{journal}{Phys. Rev. B}}
  \textbf{\bibinfo{volume}{77}}, \bibinfo{pages}{125134}
  (\bibinfo{year}{2008}).

\bibitem{Galakhov:2002hf}
\bibinfo{author}{Galakhov, V.~R.} \emph{et~al.}
\newblock \bibinfo{title}{{Mn 3$s$ exchange splitting in
  mixed-valence manganites}}.
\newblock \emph{\bibinfo{journal}{Phys. Rev. B}}
  \textbf{\bibinfo{volume}{65}}, \bibinfo{pages}{113102}
  (\bibinfo{year}{2002}).

\bibitem{Nucker:1995hk}
\bibinfo{author}{N{\"u}cker, N.} \emph{et~al.}
\newblock \bibinfo{title}{{Site-specific and doping-dependent electronic
  structure of YBa$_2$Cu$_3$O$_x$ probed by O 1s and Cu 2p x-ray-absorption
  spectroscopy}}.
\newblock \emph{\bibinfo{journal}{Phys. Rev. B}}
  \textbf{\bibinfo{volume}{51}}, \bibinfo{pages}{8529--8542}
  (\bibinfo{year}{1995}).

\bibitem{Merz:1998he}
\bibinfo{author}{Merz, M.} \emph{et~al.}
\newblock \bibinfo{title}{{Site-Specific X-Ray Absorption Spectroscopy of
  Y$_{1-x}$Ca$_x$Ba$_2$Cu$_3$O$_{7-y}$: Overdoping and Role of Apical Oxygen for High Temperature
  Superconductivity}}.
\newblock \emph{\bibinfo{journal}{Phys. Rev. Lett.}}
  \textbf{\bibinfo{volume}{80}}, \bibinfo{pages}{5192--5195}
  (\bibinfo{year}{1998}).

\bibitem{Hawthorn:2011db}
\bibinfo{author}{Hawthorn, D.~G.} \emph{et~al.}
\newblock \bibinfo{title}{{Resonant elastic soft x-ray scattering in
  oxygen-ordered YBa$_{2}$Cu$_{3}$O$_{6+\delta}$}}.
\newblock \emph{\bibinfo{journal}{Phys. Rev. B}}
  \textbf{\bibinfo{volume}{84}}, \bibinfo{pages}{075125}
  (\bibinfo{year}{2011}).

\bibitem{Zhang:1988jf}
\bibinfo{author}{Zhang, F.} \& \bibinfo{author}{Rice, T.}
\newblock \bibinfo{title}{{Effective Hamiltonian for the superconducting Cu
  oxides}}.
\newblock \emph{\bibinfo{journal}{Phys. Rev. B}}
  \textbf{\bibinfo{volume}{37}}, \bibinfo{pages}{3759--3761}
  (\bibinfo{year}{1988}).

\bibitem{Meyers:2013bf}
\bibinfo{author}{Meyers, D.} \emph{et~al.}
\newblock \bibinfo{title}{{Zhang-Rice physics and anomalous copper states in
  A-site ordered perovskites}}.
\newblock \emph{\bibinfo{journal}{Scientific Reports}}
  \textbf{\bibinfo{volume}{3}}, \bibinfo{pages}{1834} (\bibinfo{year}{2013}).

\bibitem{Peets:2009iv}
\bibinfo{author}{Peets, D.~C.} \emph{et~al.}
\newblock \bibinfo{title}{{X-Ray Absorption Spectra Reveal the Inapplicability
  of the Single-Band Hubbard Model to Overdoped Cuprate Superconductors}}.
\newblock \emph{\bibinfo{journal}{Phys.l Rev. Lett.}}
  \textbf{\bibinfo{volume}{103}} (\bibinfo{year}{2009}).

\bibitem{Chen:2013ts}
\bibinfo{author}{Chen, Y.-J.} \emph{et~al.}
\newblock \bibinfo{title}{{Doping evolution of Zhang-Rice singlet spectral
  weight: A comprehensive examination by x-ray absorption spectroscopy}}.
\newblock \emph{\bibinfo{journal}{Phys. Rev. B}}
  \textbf{\bibinfo{volume}{88}}, \bibinfo{pages}{134525}
  (\bibinfo{year}{2013}).

\bibitem{Neumeier:1989if}
\bibinfo{author}{Neumeier, J.~J.}, \bibinfo{author}{Bj{\o}rnholm, T.},
  \bibinfo{author}{Maple, M.~B.} \& \bibinfo{author}{Schuller, I.~K.}
\newblock \bibinfo{title}{{Hole filling and pair breaking by Pr ions in
  YBa$_{2}$Cu$_{3}$O$_{6.95\pm0.02}$}}.
\newblock \emph{\bibinfo{journal}{Phys. Rev. Lett.}}
  \textbf{\bibinfo{volume}{63}}, \bibinfo{pages}{2516--2519}
  (\bibinfo{year}{1989}).

\bibitem{ChNiedermayer:1998fu}
\bibinfo{author}{Niedermayer, C.} \emph{et~al.}
\newblock \bibinfo{title}{{Common Phase Diagram for Antiferromagnetism in
  La$_{2-x}$Sr$_x$CuO$_4$ and Y$_{1-x}$Ca$_x$Ba$_2$Cu$_3$O$_6$ as Seen by Muon Spin Rotation}}.
\newblock \emph{\bibinfo{journal}{Phys. Rev. Lett.}}
  \textbf{\bibinfo{volume}{80}}, \bibinfo{pages}{3843--3846}
  (\bibinfo{year}{1998}).

\bibitem{Tallon:1997gx}
\bibinfo{author}{Tallon, J.}, \bibinfo{author}{Bernhard, C.},
  \bibinfo{author}{Williams, G.} \& \bibinfo{author}{Loram, J.}
\newblock \bibinfo{title}{{Zn-induced $T_c$ Reduction in High- $T_c$ Superconductors:
  Scattering in the Presence of a Pseudogap}}.
\newblock \emph{\bibinfo{journal}{Phys. Rev. Lett.}}
  \textbf{\bibinfo{volume}{79}}, \bibinfo{pages}{5294--5297}
  (\bibinfo{year}{1997}).

\bibitem{Maiti:2009hq}
\bibinfo{author}{Maiti, K.} \emph{et~al.}
\newblock \bibinfo{title}{{Doping dependence of the chemical potential and
  surface electronic structure in YBa$_2$Cu$_3$O$_{6+x}$ and La$_{2-x}$Sr$_x$CuO$_4$ using
  hard x-ray photoemission spectroscopy}}.
\newblock \emph{\bibinfo{journal}{Phys. Rev. B}}
  \textbf{\bibinfo{volume}{80}}, \bibinfo{pages}{165132}
  (\bibinfo{year}{2009}).

\bibitem{Tanuma:2011ct}
\bibinfo{author}{Tanuma, S.}, \bibinfo{author}{Powell, C.~J.} \&
  \bibinfo{author}{Penn, D.~R.}
\newblock \bibinfo{title}{{Calculations of electron inelastic mean free paths.
  IX. Data for 41 elemental solids over the 50 eV to 30 keV range}}.
\newblock \emph{\bibinfo{journal}{Surface And Interface Analysis}}
  \textbf{\bibinfo{volume}{43}}, \bibinfo{pages}{689--713}
  (\bibinfo{year}{2011}).

\bibitem{Yang:1990bi}
\bibinfo{author}{Yang, I.-S.}, \bibinfo{author}{Schrott, A.} \&
  \bibinfo{author}{Tsuei, C.}
\newblock \bibinfo{title}{{Ba core-level shift in x-ray photoemission
  spectroscopy on single-phase Y$_{1-x}$Pr$_x$Ba$_2$Cu$_3$O$_7$ and YBa$_2$Cu$_3$O$_{7-y}$ compounds}}.
\newblock \emph{\bibinfo{journal}{Phys. Rev. B}}
  \textbf{\bibinfo{volume}{41}}, \bibinfo{pages}{8921--8926}
  (\bibinfo{year}{1990}).

\bibitem{Yang:1991jt}
\bibinfo{author}{Yang, I.-S.}, \bibinfo{author}{Schrott, A.},
  \bibinfo{author}{Tsuei, C.}, \bibinfo{author}{Burns, G.} \&
  \bibinfo{author}{Dacol, F.}
\newblock \bibinfo{title}{{Electronic states in rare-earth 1:2:3 oxides:
  Photoemission and Raman studies}}.
\newblock \emph{\bibinfo{journal}{Phys. Rev. B}}
  \textbf{\bibinfo{volume}{43}}, \bibinfo{pages}{10544--10547}
  (\bibinfo{year}{1991}).

\bibitem{Chambers:2000fo}
\bibinfo{author}{Chambers, S.~A.} \emph{et~al.}
\newblock \bibinfo{title}{{Band discontinuities at epitaxial SrTiO$_3$/Si(001) heterojunctions}}.
\newblock \emph{\bibinfo{journal}{Appl. Phys. Lett.}}
  \textbf{\bibinfo{volume}{77}}, \bibinfo{pages}{1662} (\bibinfo{year}{2000}).

\bibitem{Chambers:2004fd}
\bibinfo{author}{Chambers, S.~A.}, \bibinfo{author}{Droubay, T.},
  \bibinfo{author}{Kaspar, T.~C.} \& \bibinfo{author}{Gutowski, M.}
\newblock \bibinfo{title}{{Experimental determination of valence band maxima
  for SrTiO$_3$, TiO$_2$, and SrO and the associated valence band offsets
  with Si(001)}}.
\newblock \emph{\bibinfo{journal}{Journal Of Vacuum Science {\&} Technology B}}
  \textbf{\bibinfo{volume}{22}}, \bibinfo{pages}{2205} (\bibinfo{year}{2004}).



\bibitem{Pavlenko:2007bv}
\bibinfo{author}{Pavlenko, N.}, \bibinfo{author}{Elfimov, I.},
  \bibinfo{author}{Kopp, T.} \& \bibinfo{author}{Sawatzky, G.~A.}
\newblock \bibinfo{title}{{Interface hole doping in cuprate-titanate
  superlattices}}.
\newblock \emph{\bibinfo{journal}{Phys. Rev. B}}
  \textbf{\bibinfo{volume}{75}}, \bibinfo{pages}{140512}
  (\bibinfo{year}{2007}).

\bibitem{Pavlenko:2009kt}
\bibinfo{author}{Pavlenko, N.}
\newblock \bibinfo{title}{{Mechanism of orbital reconstruction at the
  interfaces of transition metal oxides}}.
\newblock \emph{\bibinfo{journal}{Phys. Rev. B}}
  \textbf{\bibinfo{volume}{80}} (\bibinfo{year}{2009}).

\bibitem{Salafranca:2010fga}
\bibinfo{author}{Salafranca, J.} \& \bibinfo{author}{Okamoto, S.}
\newblock \bibinfo{title}{{Unconventional Proximity Effect and Inverse
  Spin-Switch Behavior in a Model Manganite-Cuprate-Manganite Trilayer
  System}}.
\newblock \emph{\bibinfo{journal}{Phys. Rev. Lett.}}
  \textbf{\bibinfo{volume}{105}}, \bibinfo{pages}{256804}
  (\bibinfo{year}{2010}).
  
  \bibitem{yang2009}
  \bibinfo{author}{Yang, X.}   \bibinfo{author}{Yaresko, A. N.} \bibinfo{author}{Antonov, V. N.} \& \bibinfo{author}{Andersen, O.K.}
  \newblock \bibinfo{title}{{Electronic structure and x-ray magnetic circular dichroism of YBa$_2$Cu$_3$O$_7$/LaMnO$_3$ superlattices from first-principles calculations}}.
\newblock \emph{\bibinfo{journal}{arXiv:0911.4349}}
%  \textbf{\bibinfo{volume}{105}}, \bibinfo{pages}{256804}
  (\bibinfo{year}{2010}).

\bibitem{Liang:2006hr}
\bibinfo{author}{Liang, R.}, \bibinfo{author}{Bonn, D.~A.} \&
  \bibinfo{author}{Hardy, W.~N.}
\newblock \bibinfo{title}{{Evaluation of CuO$_{2}$ plane hole doping in
  YBa$_{2}$Cu$_{3}$O$_{6+x}$ single crystals}}.
\newblock \emph{\bibinfo{journal}{Physical Review B}}
  \textbf{\bibinfo{volume}{73}}, \bibinfo{pages}{180505}
  (\bibinfo{year}{2006}).

\bibitem{Salluzzo:2008dn}
\bibinfo{author}{Salluzzo, M.} \emph{et~al.}
\newblock \bibinfo{title}{{Indirect electric field doping of the CuO2 planes of
  the cuprate NdBa$_2$Cu$_3$O$_7$ superconductor}}.
\newblock \emph{\bibinfo{journal}{Phys. Rev. Lett.}}
  \textbf{\bibinfo{volume}{100}}, \bibinfo{pages}{056810}
  (\bibinfo{year}{2008}).

\bibitem{DeLuca:2010iw}
\bibinfo{author}{De~Luca, G.~M.} \emph{et~al.}
\newblock \bibinfo{title}{{Weak magnetism in insulating and superconducting
  cuprates}}.
\newblock \emph{\bibinfo{journal}{Phys. Rev. B}}
  \textbf{\bibinfo{volume}{82}}, \bibinfo{pages}{214504}
  (\bibinfo{year}{2010}).

\bibitem{GLogvenov:2009eh}
\bibinfo{author}{Logvenov, G.}, \bibinfo{author}{Gozar, A.} \&
  \bibinfo{author}{Bozovic, I.}
\newblock \bibinfo{title}{{High-Temperature Superconductivity in a Single
  Copper-Oxygen Plane}}.
\newblock \emph{\bibinfo{journal}{Science}} \textbf{\bibinfo{volume}{326}},
  \bibinfo{pages}{699} (\bibinfo{year}{2009}).

\bibitem{DiCastro:2015ux}
\bibinfo{author}{Di~Castro, D.} \emph{et~al.}
\newblock \bibinfo{title}{{High- TcSuperconductivity at the Interface between
  the CaCuO$_2$ and SrTiO$_3$ Insulating Oxides}}.
\newblock \emph{\bibinfo{journal}{Phys.Rev. Lett.}}
  \textbf{\bibinfo{volume}{115}}, \bibinfo{pages}{147001}
  (\bibinfo{year}{2015}).


\bibitem{Tabis:2014kb}
\bibinfo{author}{Tabis, W.} \emph{et~al.}
\newblock \bibinfo{title}{{Charge order and its connection with Fermi-liquid
  charge transport in a pristine high-Tc cuprate}}.
\newblock \emph{\bibinfo{journal}{Nature Communications}}
  \textbf{\bibinfo{volume}{5}}, \bibinfo{pages}{5875} (\bibinfo{year}{2014}).
  
\bibitem{Wu:2011ke}
\bibinfo{author}{Wu, T.} \emph{et~al.}
\newblock \bibinfo{title}{{Magnetic-field-induced charge-stripe order in the
  high-temperature superconductor YBa$_2$Cu$_3$O$_y$}}.
\newblock \emph{\bibinfo{journal}{Nature}} \textbf{\bibinfo{volume}{477}},
  \bibinfo{pages}{191--194} (\bibinfo{year}{2011}).

\bibitem{Chang:2012dna}
\bibinfo{author}{Chang, J.} \emph{et~al.}
\newblock \bibinfo{title}{{Direct observation of competition between
  superconductivity and charge density wave order in YBa$_2$Cu$_3$O$_{6.67}$}}.
\newblock \emph{\bibinfo{journal}{Nature Physics}}
  \textbf{\bibinfo{volume}{8}}, \bibinfo{pages}{871} (\bibinfo{year}{2012}).

\bibitem{Ghiringhelli:2012bw}
\bibinfo{author}{Ghiringhelli, G.} \emph{et~al.}
\newblock \bibinfo{title}{{Long-Range Incommensurate Charge Fluctuations in
  (Y,Nd)Ba$_2$Cu$_3$O$_{6+x}$}}.
\newblock \emph{\bibinfo{journal}{Science}} \textbf{\bibinfo{volume}{337}},
  \bibinfo{pages}{821--825} (\bibinfo{year}{2012}).

\bibitem{LeTacon:2013es}
\bibinfo{author}{Le~Tacon, M.} \emph{et~al.}
\newblock \bibinfo{title}{{Inelastic X-ray scattering in YBa$_2$Cu$_3$O$_{6.6}$ reveals
  giant phonon anomalies and elastic central peak due to charge-density-wave
  formation}}.
\newblock \emph{\bibinfo{journal}{Nature Physics}}
  \textbf{\bibinfo{volume}{10}}, \bibinfo{pages}{52--58}
  (\bibinfo{year}{2013}).

\bibitem{Comin:2015ca}
\bibinfo{author}{Comin, R.} \emph{et~al.}
\newblock \bibinfo{title}{{Symmetry of charge order in cuprates}}.
\newblock \emph{\bibinfo{journal}{Nature Materials}}
  \textbf{\bibinfo{volume}{14}}, \bibinfo{pages}{796--800}
  (\bibinfo{year}{2015}).

\bibitem{Comin:2016}
\bibinfo{author}{Comin, R.} \& \bibinfo{author}{Damascelli, A.}
\newblock \bibinfo{title}{{Resonant X-Ray Scattering Studies of Charge Order in Cuprates}}.
\newblock \emph{\bibinfo{journal}{Annu. Rev. Condens. Matter Phys.}}
  \textbf{\bibinfo{volume}{7}}, 
  \bibinfo{pages}{369-405},
  (\bibinfo{year}{2016}).

\bibitem{He2016}He J.  {\it et al.} Observation of a three-dimensional quasi-long-range electronic supermodulation in YBa$_2$Cu$_3$O$_{7-x}$/La$_{0.7}$Ca$_{0.3}$MnO$_3$ heterostructures. {\it Nature Communications} {\bf 7}, 10852 (2016).

\bibitem{Keimer:2015bn}
\bibinfo{author}{Keimer, B.}, \bibinfo{author}{Kivelson, S.~A.},
  \bibinfo{author}{Norman, M.~R.}, \bibinfo{author}{Uchida, S.} \&
  \bibinfo{author}{Zaanen, J.}
\newblock \bibinfo{title}{{From quantum matter to high-temperature
  superconductivity in copper oxides}}.
\newblock \emph{\bibinfo{journal}{Nature}} \textbf{\bibinfo{volume}{518}},
  \bibinfo{pages}{179--186} (\bibinfo{year}{2015}).

\bibitem{Tokura:2006ff}
\bibinfo{author}{Tokura, Y.}
\newblock \bibinfo{title}{{Critical features of colossal magnetoresistive
  manganites}}.
\newblock \emph{\bibinfo{journal}{Reports On Progress In Physics}}
  \textbf{\bibinfo{volume}{69}}, \bibinfo{pages}{797} (\bibinfo{year}{2006}).

\bibitem{Ricci:2013hm}
\bibinfo{author}{Ricci, A.} \emph{et~al.}
\newblock \bibinfo{title}{{Multiscale distribution of oxygen puddles in 1/8
  doped YBa$_2$Cu$_3$O$_{6.67}$}}.
\newblock \emph{\bibinfo{journal}{Scientific Reports}}
  \textbf{\bibinfo{volume}{3}} (\bibinfo{year}{2013}).

\bibitem{Campi:2015cv}
\bibinfo{author}{Campi, G.} \emph{et~al.}
\newblock \bibinfo{title}{{Inhomogeneity of charge-density-wave order and
  quenched disorder in a high-Tc superconductor}}.
\newblock \emph{\bibinfo{journal}{Nature}} \textbf{\bibinfo{volume}{525}},
  \bibinfo{pages}{359--362} (\bibinfo{year}{2015}).

\bibitem{Akahoshi:2003hk}
\bibinfo{author}{Akahoshi, D.} \emph{et~al.}
\newblock \bibinfo{title}{{Random potential effect near the bicritical region
  in perovskite manganites as revealed by comparison with the ordered
  perovskite analogs}}.
\newblock \emph{\bibinfo{journal}{Phys. Rev. Lett.}}
  \textbf{\bibinfo{volume}{90}}, \bibinfo{pages}{177203}
  (\bibinfo{year}{2003}).

\bibitem{Santos:2011tj}
\bibinfo{author}{Santos, T.~S.} \emph{et~al.}
\newblock \bibinfo{title}{{Delta Doping of Ferromagnetism in Antiferromagnetic
  Manganite Superlattices}}.
\newblock \emph{\bibinfo{journal}{Phys. Rev. Lett.}}
  \textbf{\bibinfo{volume}{107}}, \bibinfo{pages}{167202}
  (\bibinfo{year}{2011}).

\bibitem{NelsonCheeseman:2014gm}
\bibinfo{author}{Nelson-Cheeseman, B.~B.} \emph{et~al.}
\newblock \bibinfo{title}{{Polar Cation Ordering: A Route to Introducing $>$ 10\%
  Bond Strain Into Layered Oxide Films}}.
\newblock \emph{\bibinfo{journal}{Advanced Functional Materials}}
  \textbf{\bibinfo{volume}{24}}, \bibinfo{pages}{6884} (\bibinfo{year}{2014}).

\bibitem{Li:2008dy}
\bibinfo{author}{Li, Y.} \emph{et~al.}
\newblock \bibinfo{title}{{Unusual magnetic order in the pseudogap region of
  the superconductor HgBa$_2$CuO$_{4+\delta}$}}.
\newblock \emph{\bibinfo{journal}{Nature}} \textbf{\bibinfo{volume}{455}},
  \bibinfo{pages}{372--375} (\bibinfo{year}{2008}).

\end{thebibliography}
%\bibliographystyle{naturemag}

 \subsection*{Acknowledgement}   We thank Eun Ju Moon for discussions and help with transport measurements. J.W.F. would like to thank David Hawthorne for sharing the bulk YBCO XAS data. J.C. was supported by DOD-ARO under Grant No. 0402-17291 and by  the Gordon and Betty Moore Foundation's EPiQS Initiative through Grant No. GBMF4534. Work at the Advanced Photon Source, Argonne was supported by the U.S. Department of Energy, Office of Science under Grant No. DEAC02-06CH11357. This work was supported by the U.S. Department of Energy under Contract No. DE-AC02-05CH11231, and through the Laboratory Directed Research Development program, of the Lawrence Berkeley National Laboratory; and under Contract No. DE-SC0014697 through the University of California Davis. C.S.F. also acknowledges support from the APTCOM Project of the Labex Program (France) during the writing of this paper. A.X.G. acknowledges support from the U.S. Army Research Office, under Grant No. W911NF-15-1-0181 during the writing of this paper.

\end{document}